\documentclass[preprint,12pt]{elsarticle}
\usepackage{graphicx}
\usepackage{epsfig}
\usepackage{amssymb,amsopn}
\usepackage{amsmath,mathrsfs}
\usepackage{cancel}
\usepackage{mathtools}
\usepackage[vcentermath]{youngtab}
\usepackage[usenames,dvipsnames]{color}
\usepackage{url}
\usepackage{parskip}
\usepackage{comment}
\usepackage[colorlinks = true,
            linkcolor = blue,
            urlcolor  = blue,
            citecolor = blue,
            anchorcolor = blue]{hyperref}
\usepackage{tcolorbox}
\usepackage[frozencache]{minted}
\tcbuselibrary{minted,breakable,xparse,skins}

\definecolor{bg}{gray}{0.95}
\DeclareTCBListing{mintedbox}{O{}m!O{}}{%
	breakable=true,
	listing engine=minted,
	listing only,
	minted language=#2,
	minted style=default,
	minted options={%
		linenos,
		gobble=0,
		breaklines=true,
		breakafter=
		fontsize=\small,
		numbersep=8pt,
	#1},
	boxsep=0pt,
	left skip=0pt,
	right skip=0pt,
	left=25pt,
	right=0pt,
	top=3pt,
	bottom=3pt,
	arc=5pt,
	leftrule=0pt,
	rightrule=0.5pt,
	bottomrule=2pt,
	toprule=2pt,
	colback=bg,
	colframe=teal!70,
	enhanced,
	overlay={%
		\begin{tcbclipinterior}
			\fill[teal!20!white] (frame.south west) rectangle ([xshift=20pt]frame.north west);
	\end{tcbclipinterior}},
#3}
\usepackage{textpos}

\graphicspath{{Figures}}
\newcommand\pubnumber{DESY-24-045}
\def\beq{\begin{equation}}
\def\eeq#1{\label{#1}\end{equation}}
\def\eeqn{\end{equation}}

\def\beqa{\begin{eqnarray}}
\def\eeqa#1{\label{#1}\end{eqnarray}}
\def\eeqan{\end{eqnarray}}

\def\overbar#1{\overline{#1}}

\let\bar=\overbar

\def\tr{{\mbox{\rm tr}}}

\def\Dslash{\not{\hbox{\kern-4pt $D$}}}
\def\dslash{\not{\hbox{\kern-2pt $\del$}}}

\def\msb{{\bar{\ssstyle M \kern -1pt S}}}

\newcommand{\Nc}{\ensuremath{N_c}}
\newcommand{\Ng}{\ensuremath{N_g}}
\newcommand{\Nq}{\ensuremath{n_q}}

\newcommand{\Nf}{\ensuremath{n_f}}

\newcommand{\SU}{\mathrm{SU}}
\newcommand{\Y}{\ensuremath{\mathcal{Y}}}
\newcommand{\Proj}{\ensuremath{\mathbf{P}}}
\newcommand{\CG}{\ensuremath{\mathbf{C}}}
\newcommand{\Col}{\ensuremath{\mathbf{c}}}

\newcommand{\Tens}{\ensuremath{\mathbf{T}}}

\newcommand{\ttj}{\ifmmode t\bar t+\mathrm{jet} \else $t\bar t+\mathrm{jet}\,\,$\fi}

\newcommand{\C}{\mathbb{C}}

\def\msb{\ifmmode \overline{\rm MS}\,\, \else $\overline{\rm MS}\,\, $\fi}
\newcounter{bla}

\journal{Computer Physics Communications}
\begin{document}
\emergencystretch 3em
\begin{frontmatter}
	\title{Automatic generation of orthogonal multiplet bases in $\mathrm{SU}(N_c)$ color space}
\author[a]{Bakar~Chargeishvili}
\ead{bakar.chargeishvili@desy.de}

\address[a]{
  II. Institut f\"ur Theoretische Physik, Universit\"at
  Hamburg, Luruper Chaussee 149, D~--~22761 Hamburg, Germany\\
}

\begin{abstract}
  We present a software that automatically generates a multiplet color
basis for general $2\to n$ processes in quantum chromodynamics (QCD). The
construction process is guided by the decomposition of the corresponding
$\SU(\Nc)$ representation into a direct sum of irreducible
representations. The projectors of these irreducible multiplet states are
then used to construct a minimal, linearly independent basis. The
software achieves highest efficiency through a combination of the {\tt
Z3} satisfiability modulo theories solver and the symbolic manipulation
program {\tt FORM}. The resulting color basis enables the analytical
execution of involved QCD calculations that require color decomposition.
\end{abstract}
\end{frontmatter}
\begin{textblock*}{5cm}(11cm,-15cm)
  \raggedleft
  \pubnumber
\end{textblock*}
\pagebreak
% \linenumbers

% All CPiP articles must contain the following
% PROGRAM SUMMARY.

{\bf PROGRAM SUMMARY}

\begin{small}
\noindent
{\em Program Title:} {\tt OrthoBase}\\
{\em CPC Library link to program files:} (to be added by Technical Editor) \\
{\em Developer's repository link:} \url{https://github.com/rakab/OrthoBase} \\
{\em Code Ocean capsule:} (to be added by Technical Editor)\\
{\em Licensing provisions(please choose one):} BSD 3-clause \\
{\em Programming language:} Python, FORM, Z3\\
%{\em Supplementary material:}                                 \\
  % Fill in if necessary, otherwise leave out.
%{\em Journal reference of previous version:}*                  \\
  %Only required for a New Version summary, otherwise leave out.
%{\em Does the new version supersede the previous version?:}*   \\
  %Only required for a New Version summary, otherwise leave out.
%{\em Reasons for the new version:*}\\
  %Only required for a New Version summary, otherwise leave out.
%{\em Summary of revisions:}*\\
  %Only required for a New Version summary, otherwise leave out.
{\em Nature of problem:}\\
The correct description of the color content of an involved process requires a
careful analysis of the underlying symmetry properties using group theory
and an efficient manipulation of large tensorial expressions. \\
  %Describe the nature of the problem here. \\
{\em Solution method:}\\
We provide a library which uses the {\tt Z3 SMT-solver}~[1] to efficeintly perform the manipulation of Young tableaux. This is a crucial step to decompose the symmetry structure of the process. To parametrize the resulting multiplet states we follow the strategy introduced in Ref.~[2].
   \\

\end{small}
\pagebreak

\section{Introduction}
\label{sec:intro}
The improvement of the experimental accuracy at hadron colliders increases
the demand for more precise theory calculations. In practice this often requires
to account for multiple colored particles in the final state. Due to the
non-Abelian nature of quantum chromodynamics (QCD), the emission of each additional new colored particle increases
the complexity of the calculations. Moreover, to make certain calculations
possible, one is often forced to introduce a specific color basis and perform the color
decomposition. The study of the soft anomalous dimensions in the context
of resummation~\cite{Sterman:1986aj,Catani:1989ne,Chargeishvili:2022ngl} or the generation of
parton showers are the examples of such calculations~\cite{Platzer:2012np}.

\begin{figure}[h]
	\centering
	\includegraphics[width=8cm]{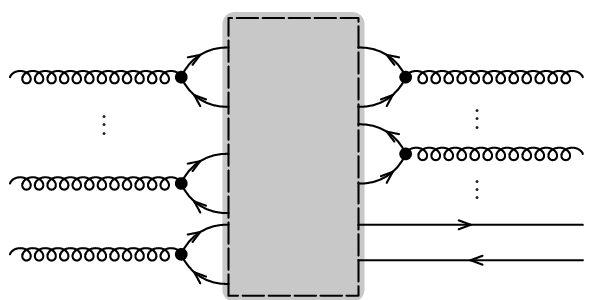}
	\caption{
		Demonstration of the construction of the trace basis. Each gluon is
		represented as a combination of quark antiquark lines. The dashed box
		should be filled with the fermion lines connecting the incoming and
		outgoing partons in all possible nonvanishing ways.
	}
	\label{fig:trace}
\end{figure}

There are a number of ways the color basis can be generated. The most widely used
one is a trace basis, which involves the generation of all possible color exchanges
between incoming and outgoing partons using the quark
lines~\cite{Cvitanovic:1976am, Mangano:1988kk}, see Figure~\ref{fig:trace}.
Later on a method better suited for Monte Carlo generators was developed based on
color flow decomposition~\cite{Maltoni:2002mq}. There are packages available
which can generate such bases~\cite{Sjodahl:2014opa,Alwall:2014hca}
automatically for any process. The issue with these kind of bases is, that they
overdescribe the color information of the process. Hence, they are not
orthogonal and this makes all the subsequent analytical calculations very
complicated.

The improved approach relies on the multiplet decomposition of the initial and
the final states and matching them with each other. This requires the
construction of appropriate projectors for each multiplet state.
The first reference of such a calculation is~\cite{MacFarlane:1968vc}, where
for ${\Nc=3}$ the color structure of a two gluon ($gg$) state has been
analyzed. The generalized method for arbitrary $\Nc$ has been formulated in
Ref.~\cite{Cvitanovic:2008zz} and it requires solving characteristic
equations of group invariant matrices. Combining the approaches of
Refs.~\cite{Cvitanovic:2008zz} and~\cite{Dokshitzer:2005ig} a general algorithm
has been formulated~\cite{Keppeler:2012ih}, which is well suited for automation
and this is the method of our choice in this work.
We also draw attention to a recently published method~\cite{Keppeler:2023msu}
which introduces a novel approach to performing color decomposition within a
multiplet basis, circumventing the explicit construction of said basis through
the utilization of Wigner 6j symbols.

The outline of the paper is as follows: In Section~\ref{sec:method} we briefly
review the method developed in Ref.~\cite{Keppeler:2012ih}. In
Section~\ref{sec:implmnt} we discuss the technical details of our
implementation. Section~\ref{sec:dmnstr} demonstrates the usage of the tools
developed in this study and Section~\ref{sec:validation} discusses the
possibilities to validate the obtained results. Finally, we conclude in
Section~\ref{sec:cnclsn} and give a future outlook.

\section{Method}
\label{sec:method}
In this Section we summarize the method formulated in~\cite{Keppeler:2012ih}:
Let us denote by $\Ng$ the number of incoming and outgoing gluons and by $\Nq$
the number of incoming quarks and outgoing antiquarks. In QCD $\Nq$ will be
also equal to the number of incoming antiquarks and outgoing quarks.
The quarks transform under the fundamental representation of $\SU(\Nc)$ and
hence, they are the elements of $V=\C^{\Nc}$, whereas antiquarks are the
elements of dual space $\overline V \cong \C^\Nc$, which transform as
complex conjugates of the fundamental representation. The gluons transform under
the adjoint representation with the dimension of $\Nc^2-1$ and they are the
elements of $A \cong \C^{\Nc^2-1}$. In this notation the color information of
every process in QCD can be characterized by a tensor
${\Col \in G \equiv (V \otimes \overline V)^{\otimes \Nq}\otimes A^{\Ng}}$.

\begin{figure}[h]
	\centering
	\includegraphics[width=0.8\linewidth]{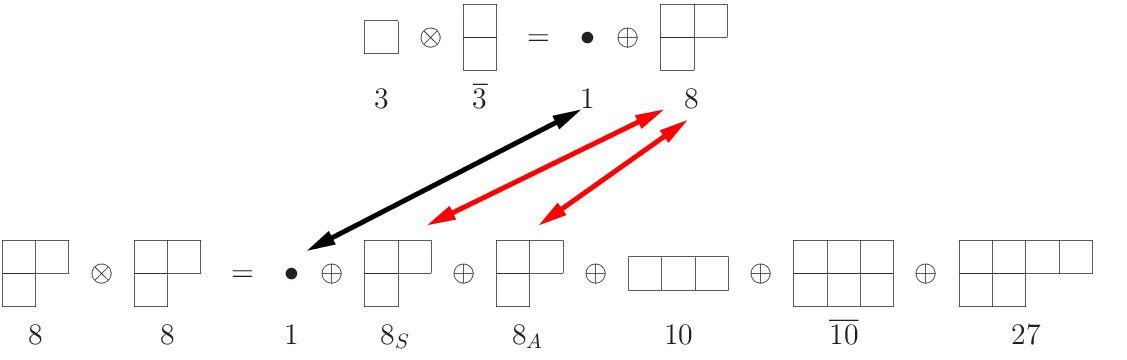}
	\caption{Multiplet decomposition and matching for $q\bar q \to gg$ in $\SU(3)$. There is one possibility to match the initial state singlet with the final state singlet and there are two possibilities to match the octets. Hence, the basis is three dimensional.} %
	\label{fig:multMatch}
\end{figure}

The idea of the construction of the multiplet color basis is to decompose $G$
as a direct sum of irreducible representations, which we refer to as
\textit{multiplets states}. Such decomposition can be achieved using the
rules of Young tableau multiplication, see, e.g., Chapter 48
in Ref.~\cite{ParticleDataGroup:2022pth}. The next step is to match the
initial state multiplets with the final state multiplets.
Figure~\ref{fig:multMatch} demonstrates this procedure for
the annihilation of a quark-antiquark pair and production of two gluons in $\SU(3)$,
i.e., $q\bar q \to gg$ scattering.

After performing of the decomposition and the matching of corresponding
multiplets, the next step is to construct projectors for each of these states.

In principle the (anti-)quark and gluon projectors require different treatments.
However, one can use the fact that for every quark-antiquark pair
$\overline{V} \otimes V = \bullet \oplus A$. The singlet state does not add
any extra color information to the system and the octet state acts as an
additional gluon. Thus, the analysis of
$(\overline{V} \otimes V)^{\otimes \Nq} \otimes A^{\otimes \Ng}$ color
structure is equivalent to the analysis of the $A^{\otimes (\Ng+\Nq)}$ color
structure. Hence, it is sufficient to focus on the generation of
gluon projectors.

The construction of the projectors depend on the creation history of the
associated multiplets. For each multiplet state the number
$\Nf(M)=0,1,2,\dotsc,\Ng$ is defined as a first appearance of multiplet $M$ in
the sequence:
\begin{equation}
  A^{\otimes 0} \, , \
  A^{\otimes 1}=A \, , \
  A^{\otimes 2} = A \otimes A \, , \
  A^{\otimes 3} = A \otimes A \otimes A\, , \
  \ldots \, , \
  A^{\otimes \Ng}.
\end{equation}

For example a $\SU(3)$ singlet has a first appearance $\Nf=0$, the octet has a
first appearance $\Nf=1$, ${\Nf(10)=\Nf(\overbar{10})=\Nf(27)=2}$ and so on.

The gluon projectors are always constructed recursively depending their
creation history.
This means, that, if the multiplet $M'$ appears in the decomposition of
$M \otimes A$, the projector for $M$ will be used for the construction.
The exact strategy depends whether the desired projector $M'$ has already
appeared previously in the sequence $A^{\otimes \Ng}$ (meaning that
$n_f(M')\leq n_f(M) < \Ng$) or whether it is an entirely new one, with $n_f=\Ng$.

First we deal with the \textit{new projectors}. This strategy can be used
to generate projectors of $10$-plet, $\overline{10}$-plet and $27$-plet
in the $\SU(3)$ expansion of $8\otimes 8$. The strategy can be summarized
as follows:
\begin{enumerate}
	\item Contract the first $\Ng-1$ gluon lines in the projector of multiplet $M$.
	\item The $\Ng$-th gluon line remains unctontracted.
	\item Each gluon line should be split into quark antiquark pairs.
	\item All the quark lines should be contracted to some Young
		operator~\cite{Cvitanovic:2008zz} $\Proj_q$ and all the antiquark
		lines should be contracted to another Young operator $\Proj_{\bar q}$.

	\item The respective Young tableaux $\Y_q$ and $\Y_{\bar q}$ should
		consist of exactly $\Ng$ cells and their product
		$\Y_{\bar q}\otimes \Y_{q}$ should contain the first occurrence of the
		multiplet $M'$.
\end{enumerate}

This procedure is illustrated below using the \textit{birdtrack} notation~\cite{Sjodahl:2014opa,Cvitanovic:2008zz}:
\begin{equation}
  \Tens^{\hdots M,M'}
  =  \parbox{9cm}{\epsfig{file=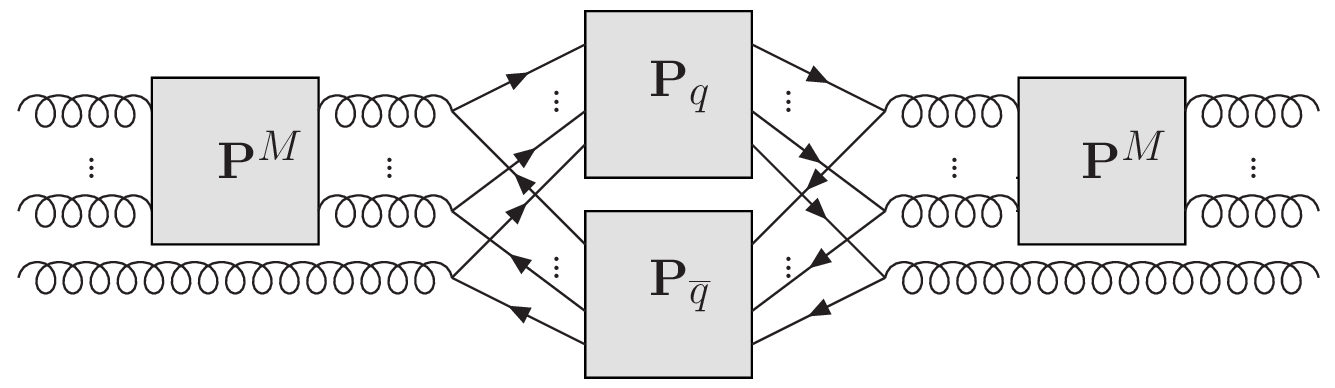,width=9cm}} \, .
  \label{eq:new_projector}
\end{equation}

Since the product $\Y_{\bar q}\otimes \Y_{q}$ will also contain known multiplets, i.e., those ones which were already previously encountered in the expansion of $A^{\otimes \Ng-1}$,
such \textit{old multiplets} have to be removed in a Gram-Schmidt step:
\begin{equation}
  \widetilde{\Tens}
  = \Tens - \sum_{M\,\text{old}} \frac{\tr(\Proj_M \Tens)}{\dim M} \, \Proj_M \, .
\end{equation}

After normalising of $\widetilde{T}$ we obtain the desired projector:
\begin{equation}
  \Proj_{M_k'} = \frac{\dim M_k'}{\tr \widetilde{\Tens}} \, \widetilde{\Tens} \, .
\end{equation}

The generation of the projectors on \textit{old multiplet} states is more
involved. This strategy is applicable for example for the projectors of
singlet and octet states in $8 \otimes 8$ expansion.
The general recipe is to construct the following operator:
\begin{equation}
  \CG^{\hdots M,M'}
  = \parbox{10cm}{\epsfig{file=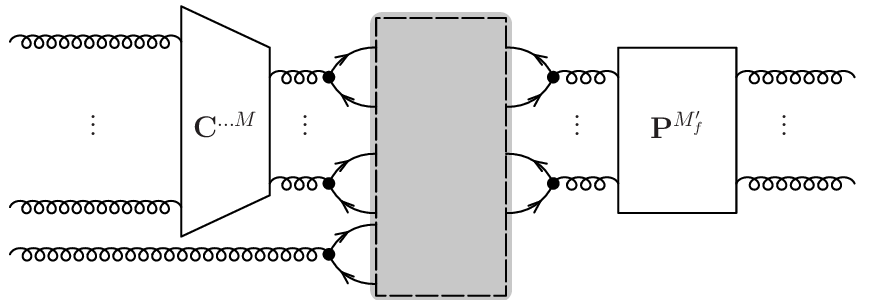,width=10cm}} \, ,
  \label{eq:any_old}
\end{equation}
where $\CG^{\hdots M}$ are Clebsch-Gordan coefficients defined as follows:
  \begin{equation}
  \label{eq:P=CCdag}
  \begin{split}
  \Proj^M \ &= \
  \parbox{5.5cm}{\epsfig{file=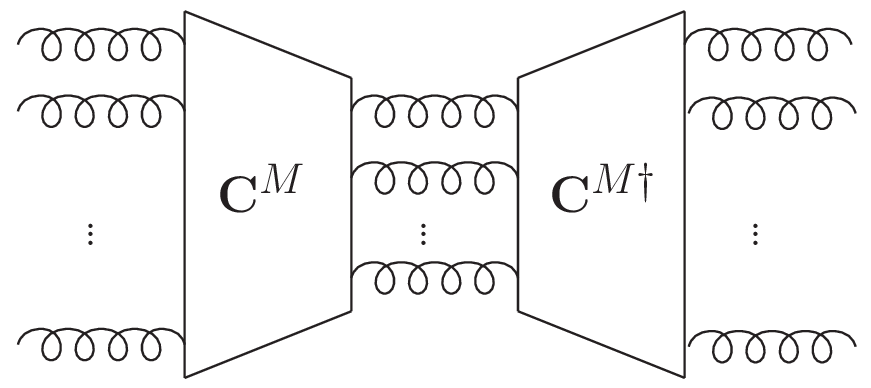,width=5.5cm}} \ .\\
  & \qquad \Ng\text{ lines} \qquad \Nf\text{ lines} \qquad \Ng\text{ lines}
  \end{split}
\end{equation}
To obtain $\CG^{\hdots M}$ appearing in the left box of Eq.~\eqref{eq:any_old}
one should apply Eq.~\eqref{eq:any_old} recursively, until everything is
expressed in terms of known Clebsch-Gordan coefficients.

In Eq.~\eqref{eq:any_old} the dashed gray rectangle connects the right quark
and antiquark lines (coming from $\Nf(M)+1$ gluons) with the left ones (recombining
in $\Nf(M')$ gluons) in such a way that the whole expression does not vanish.
After the correct normalization
$\Proj^{\hdots M,M'} = \CG^{\hdots M,M'} \CG^{\hdots M,M'\dag}$ defines the
desired projector.
In the general case, the determination of the appropriate connection inside the grey box
case can be quite cumbersome.
Hence in Ref.~\cite{Keppeler:2012ih} the special
cases were analysed, for which the procedure simplifies and this was
sufficient to analyze the $ggg \to ggg$ color structure, involving six gluons.
Below we briefly summarize these special cases, however, we also note that starting from
four gluons one will always need to turn back to general formula
in Eq.~\eqref{eq:any_old}. In the next section we will propose an
efficient algorithm to realize this formula.

There are three cases which can be simplified further:
\begin{enumerate}
	\item If $\Nf(M)<\Ng-1$ the projector is given as:
		\begin{equation}
			\Proj^{\hdots,M,M'}
			= \parbox{6cm}{\epsfig{file=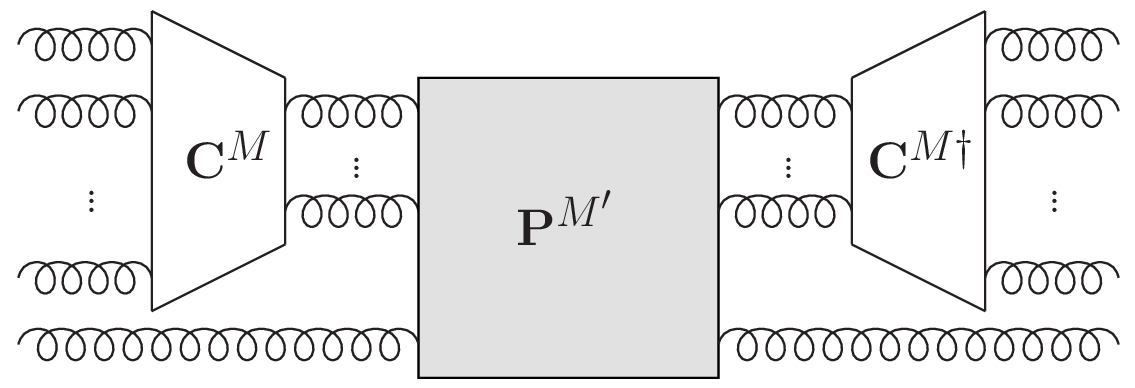,width=6cm}}
			\, .
		\end{equation}

	\item If $\Nf(M)=\Ng-1$ and $\Nf(M')=\Ng-2$:
		\begin{equation}
			\Proj^{\hdots,M,M'}
			= \frac{\dim M'}{\dim M} \
			\parbox{5cm}{\epsfig{file=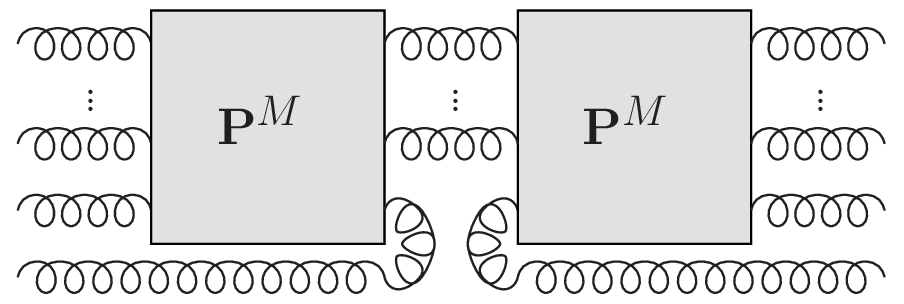,width=5cm}} \, .
			\label{eq:back_in_hist}
		\end{equation}

	\item If $\Nf(M)=\Nf(M')=\Ng-1$:
\begin{equation}
 \Proj^{\hdots,M,M'}
 = \frac{\mbox{dim}(M')}{B(M,M')}
 \parbox{7cm}{\epsfig{file=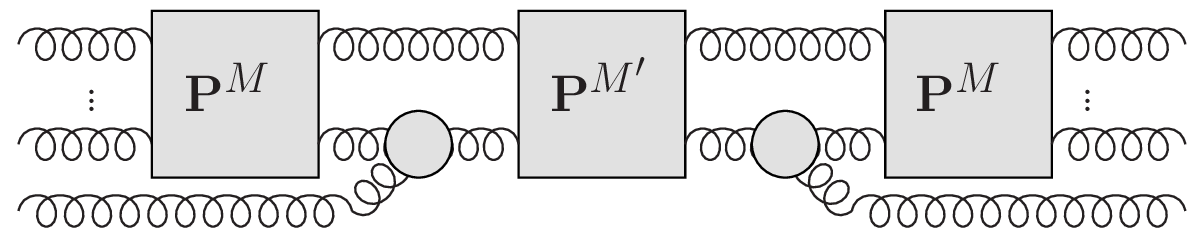,width=7cm}} \, ,
\label{eq:new_to_same_nf}
\end{equation}
where both gray circles are either $d$- or $f$-tensors or a linear combination of them and the normalization factor is determined as follows:
\begin{eqnarray}
  B(M,M')=
    \parbox{6cm}{\epsfig{file=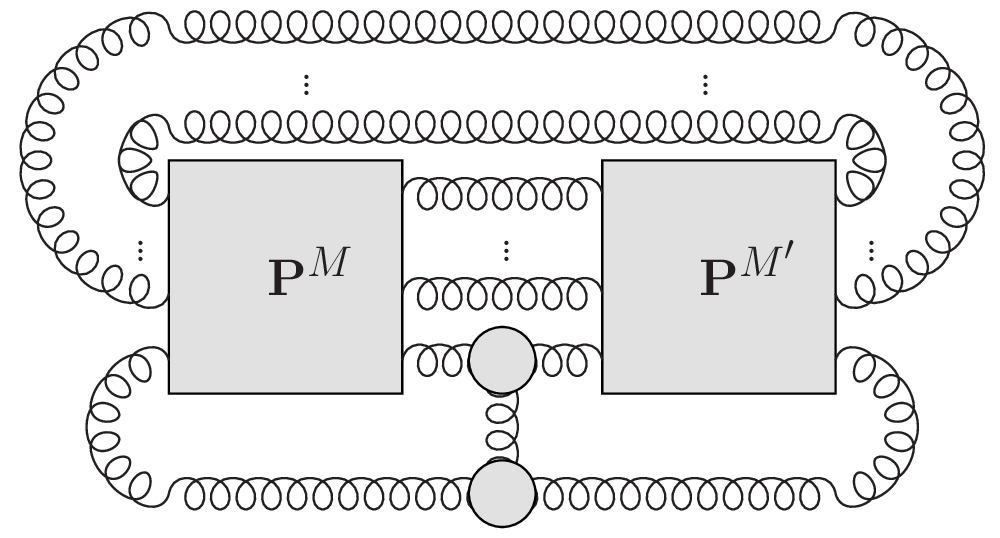,width=6cm}}
    \;.
  \label{eq:new_to_same_nfN4}
\end{eqnarray}
\end{enumerate}

For more than three gluons the above method will sometimes yield 0 and one has to refer back to Eq.~\eqref{eq:any_old}.
This finishes the summary of the ideas outlined in Ref.~\cite{Keppeler:2012ih}.
Now we describe how this is implemented in practice.

\section{Implementation}
\label{sec:implmnt}
The first step is to perform the decomposition of $A^{\otimes \Ng}$ in
terms of the irreducible representations, like it is shown on
Fig.~\ref{fig:multMatch}. This can be achieved using the rules of the
Young tableau manipulation as described
in Ref.~\cite{ParticleDataGroup:2022pth}. There are a number of {\tt
Mathematica} packages available implementing these
rules~\cite{Feger:2019tvk,Nutma:2013zea}. However, these tools only work if the
number of colors $\Nc$ is fixed in advance to some integer value, and they
become very inefficient for large values of $\Ng$. Out of the box they do not
preserve the creation history of the multiplets and the symmetry information of
the multiplets is also lost. For example, there is no distinction between $8_A$
and $8_S$ in the decomposition of $8\otimes 8$ for $\SU(3)$.

We employ a novel approach to efficiently solve these issues. The core idea of
our method is to use the {\tt Z3 SMT-solver}
\cite{10.1007/978-3-540-78800-3_24} to find the solutions of the constrained
combinatorial problems.
{\tt Z3} is a \emph{satisfiability modulo theories} (SMT) solver developed by
Microsoft Research with a purpose of solving logical formulas and constraints
over various theories, such as arithmetic, arrays, uninterpreted functions, and
bit-vectors.
In the background it employs a combination of different algorithms and
techniques to solve SMT problems efficiently. Below we list couple such
algorithms which are relevant for our problem:
{\tt Z3} is based on the \textit{Conflict Driven Clause Learning}
(CDCL)~\cite{769433}, the \textit{Davis–Putnam–Logemann–Loveland} (DPLL)
algorithm \cite{10.1145/321033.321034,10.1145/368273.368557},
the \emph{Simplex algorithm}~\cite{simplex},
the \emph{Congruence Closure algorithm}~\cite{congruence-closure},
the \emph{Fourier-Motzkin variable elimination}~\cite{fourier-motzkin},
the \emph{E-matching}~\cite{e-matching},
and the \emph{Nelson-Oppen method}~\cite{nelson-oppen}.
Hence, {\tt Z3} is superior to brute force methods because it leverages these
sophisticated algorithms and techniques that intelligently prune the search
space, handle complex theories and quantifiers efficiently, support incremental
solving, and utilize parallelization, making it highly scalable and efficient
for solving complex logical formulas and constraints, as opposed to the
computationally infeasible exhaustive exploration employed by brute force
methods.

\begin{figure}[h]
	\centering
	\renewcommand{\CancelColor}{\color{red}}
	\begin{tcolorbox}[width=0.98\textwidth]
	\scriptsize
		\begin{align*}
			\yng(2,1)\otimes\young(aa,b) &=\left( \young(\,\,a,\,)\oplus \young(\,\,,\,a)\oplus \young(\,\,,\,,a)\right)\otimes \young(a,b)\\
			                             &=\left( \young(\,\,aa,\,)\oplus \young(\,\,a,\,a)\oplus \young(\,\,a,\,,a)\oplus \cancel{\young(\,\,a,\,a)}\oplus\young(\,\,,\,a,a) \oplus \cancel{\young(\,\,a,\,a)}\oplus \cancel{\young(\,\,a,\,,a)}\right)\otimes \young(b)\\
			                             &=\cancel{\young(\,\,aab,\,)}\oplus
			                             \young(\,\,aa,\,b)\oplus
			                             \young(\,\,aa,\,,b)\oplus
			                             \cancel{\young(\,\,ab,\,a)}\oplus
			                             \young(\,\,a,\,ab)\oplus
			                             \young(\,\,a,\,a,b)\\
			                             &\oplus
			                             \cancel{\young(\,\,ab,\,,a)}\oplus
			                             \young(\,\,a,\,b,a)\oplus
			                             \cancel{\young(\,\,b,\,a,a)\oplus}
			                             \young(\,\,,\,a,ab)\\
			                             &=\young(\,\,aa,\,b)\oplus\young(\,aa)\oplus \young(\,\,a,\,ab)\oplus \young(\,a,a)\oplus\young(\,a,b)\oplus \young(\,)
		\end{align*}
	\end{tcolorbox}%
	vs.\\[0.5em]
	\begin{tcolorbox}[width=0.7\textwidth]
		\centering
		\includegraphics[width=8cm]{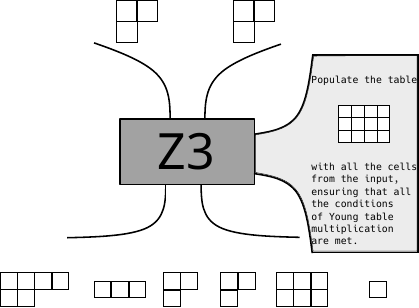}
	\end{tcolorbox}
	\caption{
		The schematic demonstration of the typical approach of
		multiplying two Young tableaux (top) vs. our
		approach using {\tt Z3 SMT-solver} (bottom).
	}
	\label{fig:old_vs_z3}
\end{figure}

When multiplying two Young tableaux, this means that instead of iteratively
adding the cells from the right tableau to the left one, we combine all the cells
together imposing the rules of the multiplication on the final tableau and let
{\tt Z3} find the solutions. Graphically this procedure is depicted on
Figure~\ref{fig:old_vs_z3}.
It is important to note that, thanks to the sophisticated algorithms listed
above, the method illustrated in the bottom scheme of Fig.~\ref{fig:old_vs_z3}
is not equivalent to generating all possible fillings of the larger tableau and
then examining which ones should be retained. Were this the case, we would
have to consider $484$ diagrams when multiplying two octets within $\SU(3)$
group.

To be able to provide the full $\SU(\Nc)$ description of the problem we
account the fact that, when multiplying $\Ng$ adjoint representations
($A^{\otimes \Ng}$) the largest possible resulting diagram can have the
dimension $\Nc\times 2\Ng$. Hence, we offer the user a possibility to carry out
the calculation for $\SU(2\Ng)$ group, which is still very efficient because of
the scalability feature of {\tt Z3} discussed above. The choice of $\SU(2\Ng)$
guarantees that every possible shape of Young diagrams will be generated. The
resulting tableaux and consequently the respective projectors are parametrized
using the $\Nc$ notation, thus providing a possibility to take the $\Nc \to 3$
limit in the end, should one be interested in applications to QCD phenomenology.

The next challenge is to construct the birdtrack diagrams described in the
previous Section. The first difficulty is to find the $\Y_q$ and $\Y_{\bar q}$
Young diagrams which are necessary for the 5th step to construct
Eq.~\eqref{eq:new_projector} for an arbitrary multiplet. Since this is a
combinatorial problem by its nature, we are using again ${\tt Z3}$ to solve it
efficiently.

Another difficulty is to find the nonzero connection inside the gray box of
Eq.~\eqref{eq:any_old}. Although this is also a combinatorial problem,
${\tt Z3}$ cannot be used directly in this case, because the restrictions
are imposed on symbolic tensorial expressions, which are not supported in
${\tt Z3}$. To construct the tensorial expressions and apply the rules of the
color algebra we use {\tt FORM}~\cite{Ruijl:2017dtg}, which we interfaced with
{\tt Z3} and let the {\tt Z3 SMT-solver} guide the writing of the ${\tt FORM}$
code. {\tt FORM} provides the result of evaluation back to {\tt Z3}. This way
we indirectly use the DPLL algorithm within ${\tt FORM}$.

The aforementioned approach  is realized in an extensible {\tt Python} library
{\tt OrthoBase}. We use a {\tt Python} \emph{Application Programming Interface}
(API) of {\tt Z3}~\cite{Bjørner2019} to implement the constraints mentioned
above. Due to the nature of the problem it is most effective if the multiplet
decomposition is performed in a serial manner. On the other hand, the projector
construction can massively benefit using parallelization. Therefore, we use
the \emph{Message Passing Interface} (MPI) protocol to not only allow
parallelization across the different cores of the CPU, but also enable
parallelization over multiple computer nodes accessible over the network. To
this end we use an open source MPI implementation
{\tt OpenMPI}~\cite{gabriel04:_open_mpi}, with a python API
{\tt mpi4py}~\cite{DALCIN20051108,9439927}.
The tensorial expressions of the resulting projectors are written as a
{\tt FORM} program, which can be embedded in other programs for further calculations.

\section{Manual of {\tt OrthoBase}}
\label{sec:dmnstr}
\subsection{Installation}
Before installing of {\tt OrthoBase}, the following dependencies need to be present
on the system: {\tt z3-solver}~\cite{z3}, {\tt Open MPI}~\cite{OpenMPI},
{\tt mpi4py}~\cite{mpi4py}. The prepackaged version of the {\tt OrthoBase} library itself can be installed
as follows:
\begin{mintedbox}{bash}
python3 -m pip install --user --upgrade OrthoBase
\end{mintedbox}
It is also possible to build the latest version from the
source of a {\tt git} repository:
\begin{mintedbox}{bash}
git clone https://github.com/rakab/OrthoBase.git
cd OrthoBase
pip install --user .
\end{mintedbox}

\subsection{Demonstartion}
We demonstrate the functionality of {\tt OrthoBase} using the illustrative
example. The code below will construct expressions of all $13\, 026\, 164$
projectors for the
${\underbracket{g g g g g g g g g g}_{\text{10 gluons}} \to {\underbracket{g g g g g g g g g g}_{\text{10 gluons}}}}$
transition for $\SU(3)$:
\begin{mintedbox}{python}
#!/usr/bin/env python3
from OrthoBase import YoungTools as YT
from OrthoBase import Projectors as P

#Define the number of colors
Nc = 3
#Define the SU(3) octet (gluon)
g = YT.YoungTableau([Nc-1,1],Nc)

#Perform the decomposition of gggggggggg state
multiplets = g*g*g*g*g*g*g*g*g*g
#Print the result of the decomposition
multiplets.print()

#Construct the projectors
projectors = P(multiplets, '/where/to/save/')
projectors.parallel_evaluation = True
projectors.nodes = [
    "MySuperComputer1.edu",
    "MySuperComputer2.edu",
    "MySuperComputer3.edu",
    ]
#Number of parallel processes for MPI
#For simple processes it is pointless to have more processes than the number of multiplets
projectors.mpi_np = 900
#Number of threads FORM is allowed to use
projectors.FORM_np = 12
projectors.run()
\end{mintedbox}

The results will be written in the directory indicated on the last line. This
path should be accessible by every node configured on line $18$. The nodes
should be able to establish the {\tt SSH}-connection with each other without
entering of the password manually (for example using {\tt SSH} keypairs) and
each of them should have same version of {\tt OpenMPI} installed along with
other dependencies of {\tt OrthoBase}. It is possible to manually configure
the options for {\tt MPI} see Section~\ref{sec:ref_manual} for more
details.

\subsection{Reference manual}
\label{sec:ref_manual}
\begin{figure}[h]
	\centering
	\includegraphics[width=14cm]{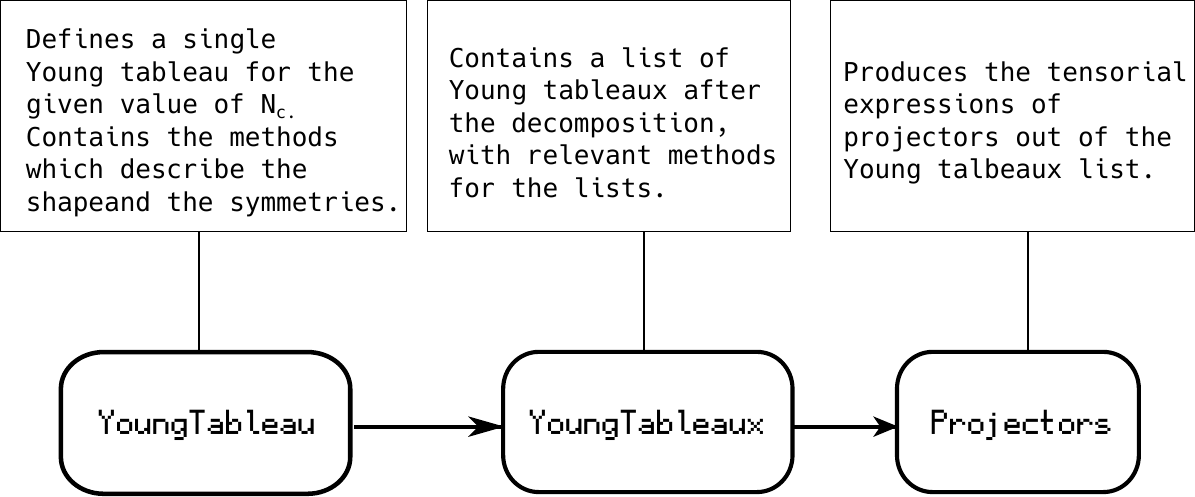}
	\caption{
		Flowchart showing the main building blocks of \texttt{OrthoBase}.
		The arrows in the flowchart represent the relationships between the
		classes: {\tt YoungTableaux} is derived from {\tt YoungTableau},
		{\tt Projectors} takes {\tt YoungTableau} as an input.
	}
	\label{fig:ortho_struct}
\end{figure}

{\tt OrthoBase} consists of two major components: \texttt{YoungTools} and {\tt Projectors}.
The relation of different classes within the library and their purposes are
summarized in Figure~\ref{fig:ortho_struct}.

Below we provide the short description of the most important functions in
these classes. The user is advised to check the online
documentation~\footnote{\url{https://orthobase.readthedocs.io/}}
which will be kept updated with the upcoming versions of the package.

\subsubsection{\tt YoungTools}
This purpose of this module is to define the object Young tableau and provide
the methods to manipulate them. This module contains two classes,
{\tt YoungTableau} and {\tt YoungTableaux}. The following program defines a
sextet and decouplet of $\SU(3)$:\\

\noindent
{\tt >>> from OrthoBase import YoungTools as YT}\\
{\tt >>> sextet = YT.YoungTableau([1,1],3)}\\
{\tt >>> decouplet = YT.YoungTableau([1,1,1],3)}\\

To decompose their direct product it is sufficient to run:\\

\noindent
{\tt >>> multiplets = sextet*decouplet}\\
{\tt >>> multiplets.print(latex=True)}\\

As a result the following output will be produced:
\begin{equation}
6\otimes 10=1\cdot \overline{15}\oplus 1\cdot 21 \oplus 1\cdot 24
\end{equation}

We note that the dimension of the multiplet does not always distinctly identify
the representation. For example the two Young tableaux below both have a
dimension $405$ in $\SU(3)$, but they clearly possess different symmetry
properties:
\begin{equation*}
	\scriptsize
	\yng(13,5)\qquad \mathrm{vs.}\qquad \yng(16,2)
\end{equation*}
In such cases the textual output of the decomposition will be marked
accordingly:
\begin{equation*}
	405_{(13,5)}\qquad \mathrm{vs.} \qquad 405_{(16,2)}
\end{equation*}

\subsubsection{\tt Projectors}
The purpose of this module is to construct the symbolic expressions of the
projectors for the multiplets contained in {\tt OrthoBase.YoungTableau}.
In addition to that, if the projectors of all previous generations are
available, the {\tt FORM}-code can be run automatically in parallel fashion
to simplify the color algebra.
If this is the case, the resulting projectors will be expressed in terms of
simple delta functions, only containing the indices of incoming and outgoing
gluons. The output will be saved in the directory provided by the user.
This module utilizes {\tt MPI} to parallelize the computation. Since the
exact choice of the parameters for {\tt MPI} depends on the particular
architechture and installation, the user is expected to provide this input.
Otherwise, the package will try to automatically determine the optimal values
and in case of the failure the whole calculation will run in the
single-threaded mode.
The online documentation provides the user access to observe the implementation
details of the configuration process.

\section{Validation}
\label{sec:validation}
There are number of ways the results can be validated. Using this package we
were able to repeat the calculation of $ggg\to ggg$ projectors and correctly
reproduce the results reported in~\cite{Keppeler:2012ih}.
For any process it is straightforward to check the orthogonality of any
pair of projectors.
Moreover, one can check the completeness relation:
\begin{equation}
	\sum_{M \in A^{\otimes \Ng}} \Proj^M_{g_1\,g_2\,\hdots,g_{\Ng},\,g_{\Ng+1}\,\hdots\,g_{2\Ng}}=\delta_{g_1\,g_{\Ng+1}}\delta_{g_2\,g_{\Ng+2}}\hdots\delta_{g_{\Ng}\,g_{2\Ng}}
	\label{eq:consis}
\end{equation}

{\tt OrthoBase} automatically generates {\tt FORM} routines for users to run
and make sure that the calculation is correct.

\section{Conclusion}
\label{sec:cnclsn}
We have presented the library \texttt{OrthoBase} which performs the
decomposition of tensorial products of an arbitrary number of $\SU(\Nc)$
representations using the rules of Young tableau manipulation and constructs the
projectors of each resulting multiplet following the method of
Ref.~\cite{Keppeler:2012ih}.
The novel approach in this context - the use of the \texttt{Z3 SMT-Solver} along with
the parallelization power of the \texttt{MPI} protocol - results in high efficiency.
The library can be used to speed up the calculations requiring the color
decomposition procedure and opens the possibility to carry out the studies for
involved processes with many colored particles. The modular design of the
library makes it possible to embed the individual parts of this calculation,
such as the manipulation of Young tableaux in other unrelated projects.
While a recently published method~\cite{Keppeler:2023msu} may prove more
suitable for certain calculations, the explicit construction of multiplet
projectors remains important. Furthermore, the combined use of an SMT-Solver
and a symbolic manipulation program opens novel avenues for calculations within
the domain of theoretical particle physics.

%\newpage
\section*{Acknowledgement}
Some of the graphs in this work were taken from Ref.~\cite{Keppeler:2012ih}, the
rest were produced using the {\tt feyn.gle} package~\cite{Grozin:2022fde}.

The author would like to thank Sven-Olaf Moch and Maria Vittoria Garzelli for
useful discussions and for valuable comments on the manuscript.\\
The author would also like to thank Daniël Boer for discussions and for cross-checking the results of the Young tableau multiplications.

\appendix
\pagebreak
\section{Examples of Young Tableaux Manipulations}
\label{apx:bases}
Print the information about some Young tableau, generate the complex
conjugate of that tableau and print the information:
\begin{mintedbox}{python}
#!/usr/bin/env python3
from OrthoBase import YoungTools as YT

y = YT.YoungTableau([2,2,2,2,2,1],3)
y.print()

yb = y.conjugate()
yb.print()
\end{mintedbox}

Iterate over the cells of the tableau, each cell is labeled with a unique
index, which can be used for later calculations (for example to construct
a Young operator):

\begin{mintedbox}{python}
#!/usr/bin/env python3
from OrthoBase import YoungTools as YT

y = YT.YoungTableau([2,2,2,2,2],3)

for pos, lab in y:
  print(f"Position: {position}, Label {lab}")
\end{mintedbox}

Find out how many adjoint representations should be multiplied to obtain the
given multiplet and how can it be represented as a quark and antiquark
tableaux:
\begin{mintedbox}{python}
#!/usr/bin/env python3
from OrthoBase import YoungTools as YT

y = YT.YoungTableau([2,2,2,2,2,2,2,1],3)
y.print()
y.decompose()

print(f"First occurrence: {y.first_occ}")
a,b,c = y.decomposition
c_dims = '-'.join(str(item.dim_txt) for item in c)
print(f"{y.dim_txt}={a.dim_txt}*{b.dim_txt}-{c_dims}"
\end{mintedbox}

Get the information about the parents of a multiplet within the
decomposition:
\begin{mintedbox}{python}
#!/usr/bin/env python3
from OrthoBase import YoungTools as YT

a = YT.YoungTableau([2,2,2,2,2,1,1],3)
b = YT.YoungTableau([2,2,2],3)
c = YT.YoungTableau([2,2,2,1],3)

for y in [a,b,c]:
    y.print()

multiplets = a*b*c

multiplets[5].parent1.print()
multiplets[5].parent2.print()

for p in multiplets[3].parent_list():
    p.print()

#Get the list of all tableaux with a dimension of 90
y_90_list = multiplets["90"]
\end{mintedbox}

Construct projectors of $gg\to gg$ with debugging logs turned on:
\begin{mintedbox}{python}
#!/usr/bin/env python3
from OrthoBase import YoungTools as YT
from OrthoBase import Projectors as P
import logging
logging.basicConfig(level=logging.WARNING)
logging.getLogger('OrthoBase').setLevel(logging.DEBUG)

#Define the number of colors
Nc = 3
#Define the SU(3) octet (gluon)
g = YT.YoungTableau([Nc-1,1],Nc)

#Perform the decomposition of gg state
multiplets = g*g
#Print information about the resulting multiplets
multiplets.print():
for m in multiplets:
  m.print()

#Construct the projectors
projectors = P(multiplets, '/where/to/save/')
#Number of threads FORM is allowed to use
projectors.FORM_np = 4
projectors.run()
\end{mintedbox}

\bibliographystyle{elsarticle-num}
\bibliography{refs}  % file
\end{document}